\documentstyle[12pt]{article}
\begin{document}

\pagestyle{plain}

\vspace*{-10mm}

\baselineskip18pt
\begin{flushright}
{\bf BROWN-HET-979}\\
{\bf BROWN- TA -521}\\
{\bf SNUTP - 94 -128}\\
{\bf hep-ph/9503xxx}\\
{\bf March 1995}\\
\end{flushright}
\vspace{1.0cm}
\begin{center}
{\Large \bf Consequences of Recent Electroweak Data and}\\
{\Large \bf W-mass for the Top Quark and Higgs Masses
\footnote{Presented
at Beyond the Standard Model IV, Granlibakken, Lake Tahoe,
CA (December 13 -19, 1994)}}\\

\vglue 5mm
{\bf Kyungsik Kang } \\
\vglue 2mm
{\it Department of Physics, Brown University
, Providence, RI, 02912 USA\footnote{
Supported
in part by the USDOE contract DE-FG02-91ER40688-Task A} }\\

\vglue 3mm
{\it and}\\
\vglue 3mm
{\bf Sin Kyu Kang } \\
\vglue 2mm

{\it Department of Physics, Seoul National University, Seoul,
Korea\footnote{Supported in part by the Basic Science Research
Institute Program, Ministry of Education, Project No. BSRI-94-2418,
and the Korea Science and Engineering Foundation through SNU CTP} }\\
\vglue 10mm
{\bf ABSTRACT} \\
\vglue 8mm
\begin{minipage}{14cm}
{\normalsize

We critically reexamine the precision tests of the standard model by coupling
the current world average value of $M_W$ with the recent LEP electroweak data
with the aid of a modified ZFITTER program to include the dominant two-loop
and QCD-EW mixed terms. The results show a clear evidence of nonvanishing
electroweak radiative corrections. The recent CDF $m_t$ is a solution of the
minimal $\chi^2$-fits to the recent LEP data set and $M_W=80.23(18)~$ GeV but
with a heavy Higgs scalar, i.e., $m_t=179~$ GeV and $m_H=300~$ GeV.
We discuss how sensitive $m_t$ and $m_H$ are depending on the exact value
of $M_W$ even within the present uncertainty, as well as on $\alpha_s$ and
$\alpha (M_Z)$. We show how the future improvements on $M_W$ can discriminate
different values of $m_t$ and $m_H$ from the electroweak data and provide a
crucial and decisive test for the standard model.
}
\end{minipage}
\end{center}
\newpage

Recent experimental advances, namely, the new
measurements of $M_W$ [1], the improved LEP precision data [2],
and the evidence of $m_t$ from CDF [3], coupled with the theoretical
progress [4,5] on the dominant two-loop and QCD-EW mixed terms, call for
a critical reexamination of the precision tests of the standard model (SM).
We would like to report on the new results of the precision tests of the SM
based on these new experimental and theoretical informations and discuss
implication on the top quark and Higgs masses as a consequence. Though global
tests of the SM with the electroweak radiative corrections (EWRC) against the
electroweak data from LEP, SLC and elsewhere have been carried
out by several group [6], the sensitivity of the tests to the exact value of
the W-boson mass [7] as well as to $\alpha_s$ and $\alpha (M_Z)$ has perhaps
not been fully recognized. For this reason we discuss in particular
how the future improvements on $M_W$, $\alpha_s$ and $\alpha (M_Z)$ can provide
a crucial test of the SM by extrapolating the consequences of the current level
of accuracy in the LEP electroweak data [2] and the new world average value of
$M_W$ [1].
Also the $m_t$ - $M_W$ correlation for different values of $m_H$ coming from
the full EWRC, when compared to the best fit solutions as well as the current
experimental values of $m_t$ and $M_W$,
reveals intriguing aspects of the precision tests for the SM and of
the prospect for new physics.  The full EWRC including the dominant two-loop
and QCD-EW mixed terms are calculated and the minimal $\chi^2$-fits to the data
are made by using a modified ZFITTER program [8] with the improved QCD
correction factor.

The basic electroweak parameters used in the numerical calculations are
the hyperfine structure constant, $\alpha  = \frac {e^2}{4\pi} =
1/137.0359895(61) $,
the four-fermion coupling constant of the $\mu$-decay,
$G_{\mu} = 1.16639(2)\times 10^{-5} ~~\mbox{GeV}^{-2}$,
and $Z$-mass which we take $M_Z = 91.1888(44) ~~\mbox{GeV}$.
Compared to the $Z$-mass,
the $W$-mass is yet to be improved, i.e., we have at best
$M_W = 80.21(16)~ $ GeV after combining the CDF measurement $M_W = 80.38(23)~ $
GeV or the {\bf new } world average value $M_W=80.23 (18)~ $ GeV [1].
Numerical computations of the full EWRC require
the mass values of the leptons, quarks, Higgs scalar, as well as $\alpha_s$
besides these quantities. The minimal $\chi^2$-fits to the data therefore can
give only correlations among $M_W$, $m_t$ within the experimental uncertainties
and $m_H$ for the given $\alpha_s$ and $\alpha (M_Z)$. The latter has a
substantial uncertainty coming from the hadronic contributions and can cause
significant shifts in the output solutions. We report here the results of the
minimal $\chi^2$-fits to the 1994 data set [2] of the Z-decay parameters
measured at LEP and to $M_W = 80.23(18)~ $ GeV.

One has, in the SM, the on-shell relation
$\sin^2 \theta_W = 1-\frac{M_W^2}{M_Z^2} $,
and the four-fermion coupling constant $G_{\mu}$
\begin{equation}
G_{\mu } = \frac{\pi \alpha }{\sqrt{2}M_W^2}
\left(1-\frac{M_W^2}{M_Z^2}\right)^{-1}\frac{1}{1-\Delta r}
\end{equation}
so that $\Delta r$, representing the radiative corrections, is
given by
\begin{equation}
\Delta r = 1-\left(\frac{A}{M_W}\right)^2
\frac{1}{1-M_W^2/M_Z^2}
\end{equation}
where $A=37.2802\pm 0.0003$.

We have found [7] that the radiative correction $\Delta r$ is
sensitive to the value of $M_W$. Mere change in $M_W$ by 0.59 $\%$ can result
as much as 75 $\%$ in $\Delta r$. Also precise determination of
the on-shell value of $\sin ^2\theta_W$
can constrain the needed value of $\Delta r$ and $M_W$. The partial width for
$Z\rightarrow f\bar{f}$ is given by
\begin{equation}
\Gamma_f = \frac{G_{\mu}}{\sqrt{2}}\frac{M_Z^3}{24\pi}\beta R_{\mbox{QED}}
c_fR_{\mbox{QCD}}(M_Z^2)\left \{ [(\bar{v}^Z_f)^2+(\bar{a}^Z_f)^2]\times
 \left(1+2\frac{m_f^2}{M_Z^2}\right)-6(\bar{a}^Z_f)^2\frac{m_f^2}
{M_Z^2}\right \}
\end{equation}
where $\beta =\beta(s)=\sqrt{1-4m_f^2/s}$ at $s=M_Z^2$,
$R_{\mbox{QED}}=1+\frac{3}{4}\frac{\alpha}{\pi}Q_f^2$,
$R_{\mbox{QCD}} = 1+1.05\frac{\bar{\alpha_s}}{\pi}+0.9(\pm 0.1)
\left(\frac{\bar{\alpha_s}}{\pi}\right)^2-13.0
\left(\frac{\bar{\alpha_s}}{\pi}\right)^3$ for the light quarks [9] and
$R_{\mbox{QCD}} = 1+c_1(m_b)\frac{\bar{\alpha_s}}{\pi}+c_2(m_b,m_t)
\left(\frac{\bar{\alpha_s}}{\pi}\right)^2-13.0
\left(\frac{\bar{\alpha_s}}{\pi}\right)^3$ for b quarks [8],
 with the gluonic coupling constant
$\bar{\alpha_s}(M_Z^2) = 0.123 \pm 0.006$ [9],
and the color factor $c_f=3$ for
quarks and 1 for leptons.
Here the renormalized vector and axial-vector couplings are defined by
$\bar{a}_f^Z=\sqrt{\rho_f^Z}2a_f^Z = \sqrt{\rho_f^Z}2I_3^f $ and
$\bar{v}^Z_f=\bar{a}^Z_f[1-4|Q_f|\sin^2\theta_W\kappa^Z_f] $ in
terms of the familiar notations [8,10].
It is customary that
all non-photonic and pure weak loop corrections in the vertices and box
diagrams are grouped in $\rho_f^Z$ and $\kappa_f^Z$ along with the propagator
corrections due to t-quark and Higgs, while all other radiative corrections
in the propagators
are contained in the couplings through $G_{\mu}$. Experimentally, the
renormalized vector and axial-vector couplings are obtained from the data
after removing all photonic contributions.

%
\begin{table}
\begin{center}
\begin{tabular}{|c||c||c|c|c|} \hline \hline
 & Experiment & Full EW & Full EW & Full EW \\ \hline
 $m_t$~(GeV) & $174\pm 10^{+13}_{-12}$ & 195 & 179 & 162 \\
 $ m_H$~(GeV) & 60 $\leq m_H \leq 1000$& 1000 & 300 & 60 \\ \hline
$M_W$~(GeV) & $80.23\pm 0.18$ & 80.39 & 80.36 & 80.35 \\
$ \Gamma_Z $~(MeV) & $2497.4\pm 3.8 $ & 2499.1 & 2499.0 & 2498.3 \\
$\sigma _h^P (nb)$ & $41.49 \pm 0.12 $ & 41.42 & 41.40 & 41.39  \\
$R(\Gamma_{had}/\Gamma_{l\bar{l}})$ & $20.795 \pm 0.040$ & 20.776 &
20.792 & 20.811 \\
$ A^{0,l}_{FB} $ & $0.0170\pm 0.0016 $ & 0.0155 & 0.0156 & 0.0159 \\
$ A_{\tau} $ & $0.143\pm 0.010 $ & 0.140 & 0.140 & 0.141 \\
$ A_{e} $ & $0.135\pm 0.011 $ & 0.140 & 0.140 & 0.141 \\
$R(\Gamma_{b\bar{b}}/\Gamma_{had})$ & $0.2202 \pm 0.0020$ & 0.2146 &
 0.2152 & 0.2158 \\
$R(\Gamma_{c\bar{c}}/\Gamma_{had})$ & $0.1583 \pm 0.0098$ & 0.1713 &
 0.1712 & 0.1711 \\
$ A^{0,b}_{FB} $ & $0.0967\pm 0.0038 $ & 0.0930 & 0.0932 & 0.0943 \\
$ A^{0,c}_{FB} $ & $0.0760\pm 0.0091 $ & 0.0596 & 0.0596 & 0.0605 \\
$ \sin^2 {\theta^{lept}_{eff}} $ from $<Q_{FB}>$ & $0.2320\pm 0.0016$ &
 0.2319 & 0.2319 & 0.2317 \\ \hline
$\chi^2 $ & & 16.5 & 14.4 & 12.1 \\ \hline
$ \Delta r $ & $0.0443\pm 0.0102$ & 0.0350 & 0.0363 & 0.0374 \\ \hline \hline
\end{tabular}
\caption{Numerical results including full EWRC for
11 experimental parameters and $M_W$.
Each pair of $m_t$ and $m_H$ represents the case of the best $\chi ^2$-
fit to the {\bf improved} 1994 LEP data [2] and $M_W = 80.23\pm 0.18$ GeV [1].}
\end{center}
\end{table}
%

The results of the best global fit to the data are given in Table 1.
One gets a stable output $M_W=80.37 \pm 0.02~$ GeV and
$m_t=179 \pm 17$ GeV for a Higgs in the range of
$m_H=60 - 1000 ~$ GeV and sees a clear effect of the EWRC.
In general the $\chi^2$-values tend to prefer
lower $m_t$ and accordingly smaller $m_H$, though any pair of ($m_t$, $m_H$)
on the Best.fit curve in Fig. 1 is statistically comparable.  The Best.fit
curve in Fig. 1
is obtained for $M_W$ = 80.23 GeV , $\alpha_s (M_Z)$ = 0.123 and $\alpha (M_Z)$
= 1/128.786. Fig. 2 shows how $M_W$ changes with $m_t$ for a fixed
$m_H$ from the full EWRC, along with the minimal $\chi^2$-fit solutions
($\Diamond$ points) as well as the world average $M_W$ and CDF $m_t$
for comparison. However the Best.fit curve in Fig. 1 and the $\Diamond$ points
in Fig. 2 can have as much as $\pm 6$ GeV and $\pm 40$ MeV
shifts in $m_t$ and $M_W$ respectively due to the uncertainty $\Delta \alpha_s
(M_Z) = \pm 0.006~$. Also there can be additional downward shifts
by 5 GeV and 20 MeV respectively upon $\alpha (M_Z)$ decreasing to
1/128.855. Note from Fig. 2 that a higher $M_W$ is preferred for a lighter
Higgs
but to distinguish a shift of 200 GeV in $m_H$ at $M_W$ = 80.23 GeV one will
need an improvements of about 50 MeV for the W-mass, i.e., better than the
theoretical uncertainty of the current precision tests.
This ambiguity in $M_W$ is about the level of the accuracy
aimed at LEP-200  and therefore the expected $m_t$ improvement at LEP-200 will
be at the best of the order 5 GeV.
We see from Fig. 1
that for a top quark not exceeding $200~$ GeV the upper bound of
$m_H$ is 300(500) GeV at $95\%$ ($90\%$) confidence level.
Fig. 2 shows
that the central values of the world average $M_W$
and CDF $m_t$ are consistent with a Higgs scalar mass somewhat heavier
than 1000 GeV to be contrasted to our output solution of the global fit.
Even with the mass dependent QCD factor, we see that there is still 2.5
$\sigma$ deviation in $R(\Gamma_{b \bar{b}}/\Gamma_{had})$ from experiment
irrespective to the uncertainty in $\alpha_s$ [5], which  may be due to
new physics beyond the SM.

In short, we find definite support for the
evidence of the nonvanishing weak-loop correction from the current world
average $M_W$ and LEP data. In particular,
the CDF $m_t$ is a solution of the minimal $\chi^2$-fit to
the current LEP data and the world average value of $M_W$ but
with a Higgs $300 \pm 200~$ GeV depending on the
input value of $\alpha_s$.
Thus, improved measurements of $M_W$ within 50 MeV accuracy
in the future precision experiments
can provide a crucial test of the SM as it will start to
distinguish different Higgs mass
to within 200 GeV.

\section*{References}
\begin{description}
\item[1.] D. Saltzberg, Fermilab-Conf-93/355-E; C.K. Jung, in:Proc. 27th ICHEP
          (Glasgow, July 1994).
\item[2.] P. Clarke, Y. K. Kim, B. Pictrzyk, P. Siegris, and M. Woods, in:
          Proc. 29th Rencontres de Moriond (Meribel,1994); R. Miquel, in:
          Proc. 22nd INS Symp. (Tokyo, March 1994); D. Schaile, in:
	  Proc. 27th ICHEP (Glasgow, July 1994).
\item[3.] F. Abe et al., Phys. Rev.{\bf D 50} (1994) 2966; Phys.Rev.Lett.
          {\bf 73} (1994) 255; S. Abachi et al, Phys. Rev. Lett. {\bf 72}
          (1994) 2138; FERMILAB-PUB-94/354-E.
          {\bf Note added}: The most recent $m_t$ value from the CDF is
          $m_t = 176 \pm 8 \pm 10 $ GeV, while from D0 is
          $m_t = 199^{+19}_{-21} \pm 22$ GeV as reported in F. Abe etal.,
          FERMILAB-PUB-95/022-E and S. Abachi et al., FERMILAB-PUB-95/028-E.
\item[4.] B.A. Kniehl, KEK-TH-412 (1994).
\item[5.] S. Franchiotti, B. Kniehl and A. Sirlin, Phys. Rev. {\bf D 48}
          (1993) 307; B.H. Smith and M.B. Voloshin, UMN-TH-1241/94.
\item[6.] K. Hagiwara, S. Matsumoto, D. Hait and C.S. Kim, KEK preprint
          93-159 (to appear in Z. Phys. C); J. Erler and P. Langacker,
          UPR-0632T (October, 1994). For the earlier works, see
          F.Dydak, in: Proc. 25 Int. Conf. H. E. Phys., Eds. K.Phua,
          Y.Tamaguchi (World Scientific, Singapore, 1991); W. Hollik, in:
          {\it Precision Tests of the Standard Model}, ed. P.
          Langacker (World Scientific Pub., 1993); and G. Altarelli, in: Proc.
          Int.EPS. Conf. H. E. Phys. (Marseille, July 1993).
\item[7.] Kyungsik Kang and Sin Kyu Kang, Brown HET-940 (March 1994);
          SNUTP-94-59 (June 1994); Brown HET-968 (in: Proc. Workshop
          on Quantum Infrared Physics, Paris, June 1994). K. Kang,
          Brown-HET-931 (December 1993) (in: Proc. 14th Int. Workshop
          Weak Interactions and Neutrinos, Seoul, July 1993); Z. Hioki,
          Tokushima preprint 94-01; TUM-T31-80/94 (October 1994).
\item[8.] D. Bardin et al., CERN-TH-6443-92 (1992).
\item[9.] T. Hebbeker, Aachen preprint PITHA 91-08 (1991);
          S.G. Gorishny, A.L. Kataev and S.A. Larin, Phys.Lett. {\bf B 259}
          (1991) 144; L.R. Surguladze and M.A. Samuel,
          Phys.Rev.Lett. {\bf 66} (1991) 560.
\item[10.] W. Hollik (Ref. 6); CERN Yellow Book CERN 89-08,
	  vol.1, p45; K. Kang (Ref. 7).

\section*{Figure Captions}
\item[Fig. 1]: The mass ranges of $m_t$ and $m_H$ from the minimal $\chi^2$-fit
to the 1994 LEP data and $M_W$ = 80.23 GeV
\item[Fig. 2]: $M_W$ versus $m_t$ for fixed values of $m_H$ from the full
 radiative correction in the standard model.
 The case of the minimal $\chi^2$-fit to the 1994 LEP data with
 the full EWRC in Table 1 are indicated by $\Diamond $.

\end{description}
\end{document}